\documentclass[superscriptaddress,twocolumn,secnumarabic,
 amssymb,amsmath,nobibnotes,aps,prd,showkeys,showpacs]{revtex4}

\usepackage{graphicx}
\usepackage{bm}%

\begin{document}


\title{Dynamics and chaos in the unified scalar field Cosmology}

\author{Georgios Lukes-Gerakopoulos}
\affiliation{Academy of Athens, Research Center for Astronomy,
 Soranou Efesiou 4, GR-11527, Athens, GREECE}
\affiliation{University of Athens,Department of Physics, Section of Astrophysics,
 Astronomy and Mechanics}

\author{Spyros Basilakos}
\affiliation{Academy of Athens, Research Center for Astronomy,
 Soranou Efesiou 4, GR-11527, Athens, GREECE}

\author{George Contopoulos}
\affiliation{Academy of Athens, Research Center for Astronomy,
 Soranou Efesiou 4, GR-11527, Athens, GREECE}

\begin{abstract}
We study the dynamics of the closed scalar field FRW cosmological models in the 
framework of the so called {\it Unified Dark Matter} (UDM) scenario. Performing a theoretical
as well as a numerical analysis we find that there is a strong indication of chaos in 
agreement with previous studies. We find that a positive value of the spatial curvature is 
essential for the appearance of chaoticity, though the Lyapunov number seems to be 
independent of the curvature value. Models that are close to flat
($k \rightarrow 0^{+}$) exhibit a chaotic behavior after a long time while pure flat 
models do not exhibit any chaos. Moreover, we find that some of the 
semiflat models in the UDM scenario 
exhibit similar dynamical behavior with the $\Lambda$ cosmology despite their chaoticity. 

Finally, we compare the measured evolution of the Hubble 
parameter derived from the differential ages of passively evolving galaxies
with that expected in the semiflat unified scalar field cosmology. 
Based on a specific set of initial conditions we find 
that the UDM scalar field model matches well the observational data.

\end{abstract}
\maketitle

\section{Introduction}

There is by now convincing evidence that the available high quality cosmological 
data (Type Ia supernovae, CMB, etc.) are well fitted by an emerging cosmological 
model, which contains cold dark matter to explain clustering and an extra 
component with negative pressure, the vacuum energy (or in a more general setting
the ``dark energy''), to explain the observed accelerated cosmic expansion 
(Refs. \cite{Riess07,Spergel07} and references therein).
Due to the absence of a physically well-motivated fundamental theory, there have 
been many theoretical speculations regarding the nature of the above exotic dark energy. 
Indeed, many authors claim that a real scalar field which rolls down 
the potential $U(\phi)$ \cite{Ozer87,Peebles88,Weinberg89,Turner97,Caldwell98,Varum00,Padm03,Peebles03}
could resemble the dark energy.

On the other hand, the global dynamics, in models with dark energy, become more complicated
than in matter dominated models. Within this framework, a serious issue is the 
existence (or nonexistence) of chaos in the scalar field cosmology. This is very important 
because chaotic fields can provide possible solutions to the cosmological coincidence problem
(the question why $\Omega_m$ and $\Omega_\Lambda$ are of the same order at the present time) 
as well as to the problem of uniqueness of vacua. The first paper on chaos in 
a closed Friedman Robertson Walker geometry coupled to a scalar field 
(FRW with a scalar field potential $U(\phi)\propto \phi^{2}$) was written by
Page \cite{Page84}. This author, suggested that the dynamical behavior of this non-integrable model is 
similar to the chaos appearing in ergodic systems. Since then, the problem of chaos in 
FRW scalar field cosmologies has been investigated thoroughly in several papers
\cite{Calzetta93,Cornish96,Kamen97,Peebles03,Joras03,Kanekar01,Beck04,Heinz05,Szydlow05,Toporensky06,Szydlow07}.
The dynamical properties of such cosmological models are met in dynamical 
systems exhibiting chaotic
scattering. However, in this kind of studies there is a crucial question 
which has to be addressed:
{\it how can we estimate the amount of chaos?} It has been shown that in 
chaotic scattering problems
classical tools, like the Lyapunov characteristic number, are unable 
to provide a clear answer whether
an orbit is chaotic or not \cite{Contop99}.

The aim of this work is to investigate the dynamical properties of a family of cosmological models
based on a scalar field potential which appears to act both as a dark matter and dark energy 
\cite{Varum00,Bertacca06}. In order to measure the amount of 
chaos we follow the notation of Ref.\cite{Contop99}
that was able to give useful answers to their scattering problem. These authors used an index 
of chaoticity which is capable to detect stochastic areas in scattering problems. 
To this end, we attempt to compare our theoretical predictions with 
observational data by utilizing the measured evolution of the Hubble 
parameter \cite{Simon05}.

The structure of the paper is as follows. The basic theoretical elements of the problem are 
presented in section 2. In particular, in the first subsection we study the dynamical system 
analytically and in the second subsection we consider various chaoticity indices. Section 3 
outlines the numerical results of the present study together with a thorough discussion, and 
finally we draw our conclusions in section 4.

\section{Theoretical elements}

\subsection{Robertson-Walker cosmology coupled with a scalar field.}

Observationally, the universe at large scales appears to be in general homogeneous and isotropic.
Such a universe can be described by the Robertson-Walker (RW) line element:
\begin{equation}
      ds^2=-dt^2+\alpha^2(t) \frac{d^2 r+r^2 d^2 \theta + r^2 
      \sin^2{\theta} d^2 \phi}{(1+\frac{k r^2}{4})^2}
      \label{RWmetric}
\end{equation}
where $\alpha(t)$ is the scale factor of the universe and the sign of $ k $ determines the 
geometry of the space. The RW cosmology has been supplemented in recent years by the theories
of inflation, dark matter and dark energy. Many of these complementary theories are based 
on the presence of one or more scalar fields. In this work we have ignored both the coupling
of the scalar field $\phi$ to other fields and quantum-mechanical effects. Doing so, it is 
straightforward to associate the gravitational RW universe with the scalar field by utilizing
the so called stress-energy tensor (see Ref. \cite{Turner83}):
\begin{eqnarray}
      T^{0}_{~0} &=& -\left[\frac{1}{2} \dot{\phi}^2 + U(\phi)\right]  \nonumber \\
      T^{i}_{~j} &=& \frac{1}{2} \dot{\phi}^2 - U(\phi) ~~~~~~~ for ~ i=j \\
      T^{\mu}_{\nu} &=& 0 ~~~~~~~~~~~~~~~~~~~~ for ~ \mu \neq \nu \nonumber
      \label{SFTensor}
\end{eqnarray}
where the overdot denotes derivatives with respect to time, $U(\phi)$ is the potential
of the scalar field, the Latin indices indicate the space coordinates, while the Greek 
indices the whole spacetime.

Using now the Einstein's field equations
\footnote {In this work we set the speed of light $c \equiv 1$ and  $ 8 \pi G \equiv 1 $.}
\begin{equation}
      R_{\mu\nu}-\frac{1}{2} g_{\mu\nu}R=T_{\mu\nu}
      \label{EFEint}
\end{equation}
and the equation of state through conservation of the stress-energy:
\begin{equation}
      T_{\mu \nu ~; \mu}=0.
      \label{CoSE}
\end{equation}
the corresponding Friedmann equations become:
\begin{eqnarray}
     3 \left[ \left( \frac {\dot{\alpha}} {\alpha} \right )^2+\frac{k}{\alpha^2} \right] &=& 
     \frac{\dot{\phi}^2}{2}+U(\phi) \label{EFE1} \\
     2 \left(\frac{\ddot{\alpha}}{\alpha}\right)+\left(\frac {\dot{\alpha}} {\alpha} 
     \right)^2+\frac{k}{\alpha^2} &=& -\frac{\dot{\phi}^2}{2}+U(\phi)
     \label{EFE2}
\end{eqnarray}
while the equation of motion for the scalar field takes the form:
\begin{equation}
      \ddot{\phi}+3 \frac{\dot{\alpha}}{\alpha}\dot{\phi}+\frac{\partial{U}}{\partial{\phi}}=0\;\;.
      \label{SEC}
\end{equation}

The above set of differential equations describes the evolution of the Robertson-Walker cosmology 
when gravity is coupled with a scalar field. It can be proved (see Ref. \cite{Page84,Gousheh00})
that these equations of motion can also emerge from the Lagrangian:
\begin{equation}
      L=-3\alpha \dot{\alpha}^2 + 3 k \alpha + \alpha^3 \left[\frac{\dot{\phi}^2}{2}-U(\phi)\right] \;\;.
      \label{Langaf}
\end{equation}
A serious problem that hampers the straightforward use of such an approach is our ignorance
of the form of the potential $ U(\phi)$. Note that in the literature, because of the absence of a physically
well-motivated fundamental theory, there are plenty of such potentials (for a review see \cite{Toporensky06}) 
which approach differently the nature of the scalar field. For example, if the main term in the potential
$ U(\phi)$ is $\phi^{n}$ then the energy density of the scalar field is $\rho_{\phi} \propto \alpha^{-6n/(n+2)}$.
Therefore, it becomes evident that for $n=2$ or $n=4$, the corresponding energy density behaves either like non 
relativistic or relativistic matter \cite{Turner83}.

In a recent work \cite{Bertacca06}, the authors used the so called {\it Unified Dark Matter} (UDM) scenario
in order to parameterize the functional form of the potential
\begin{equation}
      U(\phi) =c_{1} \cosh^2{(c~\phi)}+c_2   ~~~~~  c, c_1,~ c_2 ~ \in ~ \Re
      \label{potent} \;\;.
\end{equation}
From a cosmological point of view, this ''cosmic'' fluid behaves both as a dark energy and dark matter.
However, at an early enough epoch the fluid evolves like radiation \cite{Scherrer04}.
Also it has been found \cite{Bertacca06} that in a spatially flat RW model ($k=0$)
the UDM fluid predicts the same global dynamics as the $\Lambda$ cosmology does. 

The real constants $c_1$ and $ c_2 $ in Eq. (\ref{potent}) obey certain restrictions because the square
 of the mass of the scalar field is equal to $ \frac{\partial^2 {U(\phi)}}{\partial{\phi}^2} \geq 0 $
for $ \phi=0 $ and furthermore, according to Ref. \cite{Bertacca06}, $ U(0) \geq 0 $, otherwise the scalar
field mass could get negative values. Explicitly these restrictions mean for $ c_1 $ and $ c_2 $ that:
\begin{eqnarray}
      c_1 &\geq& 0 \label{potcon1} \\
      c_1 &\geq& -c_2 \label{potcon2}
\end{eqnarray}
Changing variables from $(\alpha,\phi)$ to $(x,y)$ using the relations
\begin{eqnarray} 
      \label{trans}
      x=A ~\alpha^{3/2} \sinh{(c ~\phi)} \nonumber \\
      y=A ~\alpha^{3/2} \cosh{(c ~\phi)}
\end{eqnarray}
with $ A^{2}=1/c^{2} =8/3$ the Lagrangian (\ref{Langaf}) is written:
\begin{eqnarray}
      \label{Langxy}
      L=\frac{1}{2}\left[(\dot{x}^2+\frac{3}{4}c_2~ x^2)-
       (\dot{y}^2+\frac{3}{4}(c_1+c_2)~ y^2)\right]  \nonumber \\
        +\frac{1}{2}\left[3^{4/3} k (y^2-x^2)^{1/3}\right]
\end{eqnarray}
Hence, it is obvious that in the new coordinate system our problem is described by two coupled oscillators.
Notice that the oscillator on the $y$ axis is hyperbolic due to the restriction (\ref{potcon2}), 
but the oscillator on the $x$ axis is either hyperbolic, if $ c_2 \geq 0 $, or elliptic, if $ c_2 < 0$.
Also note that $x$ and $y$ are constrained due to Eq. (\ref{trans}) which gives:
\begin{equation}
      y \ge 0, ~y \ge |x| 
      \label{varcon}
\end{equation}
because the scale factor
\begin{equation}
      \alpha = \left( \frac{y^2-x^2}{A^2}\right) ^{1/3}
      \label{alcon}
\end{equation}
is physically meaningful only if it is greater or equal to $ 0 $.

As can be seen from the above analysis there are two cases of interest:
\begin{enumerate}
      \item $ c_2 \geq 0 $
      \item $ c_2 < 0 $
\end{enumerate}
and we investigate them in the subsequent subsections. Finally, it is obvious that for a flat 
cosmological model ($k=0$) the dynamical system is fully integrable.

\subsubsection{Case $ c_2 \geq 0 $}
In this case, as mentioned before, both oscillators are hyperbolic. From the Lagrangian 
(\ref{Langxy}) we find the Hamiltonian:
\begin{eqnarray}
      \label{Hamxy1}
      {\cal H} =\frac{1}{2}\left[(p_{x}^2-\omega_1^2 x^2)-(p_{y}^2-\omega_2^2 y^2)\right] \nonumber \\
       -\frac{1}{2}\left[3^{4/3} k (y^2-x^2)^{1/3}\right]
\end{eqnarray}
where $ p_x=\dot{x} $, $ p_y=-\dot{y}$ denote the canonical momenta and 
$ \omega_1^2 = \frac{3}{4}c_2 $, $ \omega_2^2 = \frac{3}{4}(c_1+c_2) $ are the oscillators'
frequencies.

The Hamiltonian (\ref{Hamxy1}) gives the following equations of motion:
\begin{eqnarray}
      \label{eqmot1}
      \dot{x}=\frac{\partial{\cal H}}{\partial{p_x}} &=& p_x \nonumber \\
      \dot{y}=\frac{\partial{\cal H}}{\partial{p_y}} &=& -p_y \nonumber \\
      \dot{p_x}=-\frac{\partial{\cal H}}{\partial{x}} &=& 
      -\left[-\omega_1^2+\frac{3^{1/3} k}{(y^2-x^2)^{2/3}} \right] x \\
      \dot{p_y}=-\frac{\partial{\cal H}}{\partial{y}} &=& 
      \left[-\omega_2^2+\frac{3^{1/3} k}{(y^2-x^2)^{2/3}} \right] y 
      \nonumber 
\end{eqnarray}
From Eqs. (\ref{eqmot1}) and the restrictions (\ref{varcon}) we find that the 
equilibrium point of the system is:
\begin{equation}
      y = \frac{3^{1/4}~ k^{3/4}} {\omega_2^{3/2}},~~~~p_y=p_x=x=0 \label{eqp1a} 
\end{equation}
when $ k>0 $. If $ k \leq 0 $, then there are no equilibrium points which obey the restrictions (\ref{varcon}).

\subsubsection{Case $ c_2 < 0 $}

In this case we get one hyperbolic and one elliptic oscillator and the Hamiltonian becomes:
\begin{eqnarray}
      \label{Hamxy2}
      {\cal H}=\frac{1}{2}\left[(p_{x}^2+\omega_1^2 x^2)-(p_{y}^2-\omega_2^2 y^2)\right] \nonumber \\
       -\frac{1}{2}\left[3^{4/3} k (y^2-x^2)^{1/3}\right]
\end{eqnarray}
where $ \omega_1^2 = -\frac{3}{4}c_2 $.

The Hamiltonian (\ref{Hamxy2}) now provides a somewhat different set of equations than 
before:
\begin{eqnarray}
      \label{eqmot2}
      \dot{x}=\frac{\partial{\cal H}}{\partial{p_x}} &=& p_x \nonumber \\
      \dot{y}=\frac{\partial{\cal H}}{\partial{p_y}} &=& -p_y \nonumber \\
      \dot{p_x}=-\frac{\partial{\cal H}}{\partial{x}} &=& 
      -\left[\omega_1^2+\frac{3^{1/3} k}{(y^2-x^2)^{2/3}} \right] x \\
      \dot{p_y}=-\frac{\partial{\cal H}}{\partial{y}} &=& 
      \left[-\omega_2^2+\frac{3^{1/3} k}{(y^2-x^2)^{2/3}} \right] y 
      \nonumber 
\end{eqnarray}
From Eqs. (\ref{eqmot2}), and by taking into account the restriction (\ref{varcon}),
we find that the equilibrium point of the system is again:
\begin{equation}
      y = \frac{3^{1/4}~ k^{3/4}} {\omega_2^{3/2}},~~~~p_y=p_x=x=0 \label{eqp2a}
\end{equation}
when $ k>0 $, which is the same as in the previous paragraph (see Eq. (\ref{eqp1a})). 
If $ k \leq 0 $ then there are again no equilibrium points. The physical interpretation of the 
non existence of equilibrium points when the universe is flat or has negative 
curvature is that the universe cannot collapse therefore there is no other 
flow in the phase space than the one which starts with $ \alpha = 0 $ 
and tends to $ \alpha \rightarrow \infty $.

For both cases of $ c_2 $ at the equilibrium point the eigenvalues and
the corresponding eigenvectors are:
\begin{eqnarray}
\label{eig1}
 \pm\sqrt{-\frac{3}{4} c_1} &~& (p_x,p_y,x,y)=(\pm\sqrt{-\frac{3}{4} c_1},0,1,0) \\
 \label{eig2} \pm\sqrt{c_1+c_2} &~& (p_x,p_y,x,y)=(0,\pm\sqrt{c_1+c_2},0,1) 
\end{eqnarray}
Due to the restriction (\ref{potcon1}) the eigenvalues of (\ref{eig1}) are imaginary,
but due to the restriction (\ref{potcon2}) the eigenvalues of (\ref{eig2}) are
real, so we have a saddle-center instability. The saddle shape is displayed on
the $y$, $p_y$ plane. Notice that, although the existence and the position of
the equilibrium point depend from the curvature's value, the eigenvalues and the
corresponding eigenvectors for the equilibrium point are independent from the curvature.

\subsection{Chaotic indicators and scattering}
The detection of chaos  (see \cite{Contop02} for a review) is a topic for which 
many tools have been developed. These methods can be separated in two main categories. 
Tools which exploit the properties of the deviation vector and tools that are based
upon a frequency analysis. In this paper we use tools of the first category.
The deviation vector $ \xi $ is defined as the solution of the variational 
equations of the system.

A classical tool for the detection of chaos is the Lyapunov Characteristic 
Number (LCN):
\begin{equation}
      \label{LCN}
      LCN = \lim_{t \rightarrow \infty} \chi(t),
\end{equation}
where:
\begin{equation}
      \label{FLCN}
      \chi(t) = \frac{1}{t} \ln{\frac{\xi(t)}{\xi(0)}}
\end{equation}
is the ``finite time LCN''. The orbits with $LCN>0$ are called chaotic, while
the orbits with LCN=0 are called ordered. A similar useful quantity is the stretching
number \cite{Voglis94}, also called the Lyapunov indicator
\cite{Froes93}:
\begin{equation}
      \label{strnum}
      sn(t) = \ln{\frac{\xi(t+dt)}{\xi(t)}}
\end{equation}
In scattering problems when orbits escape from the region where they behave stochastically,
and then get far away from this region, they diverge from nearby orbits linearly. Therefore the 
LCN is zero. The latter behavior is also a characteristic of regular orbits. 
Hence if we entrust entirely the LCN in scattering problems we may 
characterize as regular those orbits, that in a certain region exhibit stochastic behavior.
However in our paper, in order to study chaos we use also the evolution of the stretching number,
as done in Ref. \cite{Contop99}. This method is based on the idea that when an orbit in a chaotic
scattering problem passes through a stochastic region of the phase space the average value of 
stretching varies around a positive number, even if afterwards it tends to a value near zero. 

\section{Numerical results. Comparison with observational data}

 \begin{figure*}
         \centerline{\includegraphics[width=35pc] {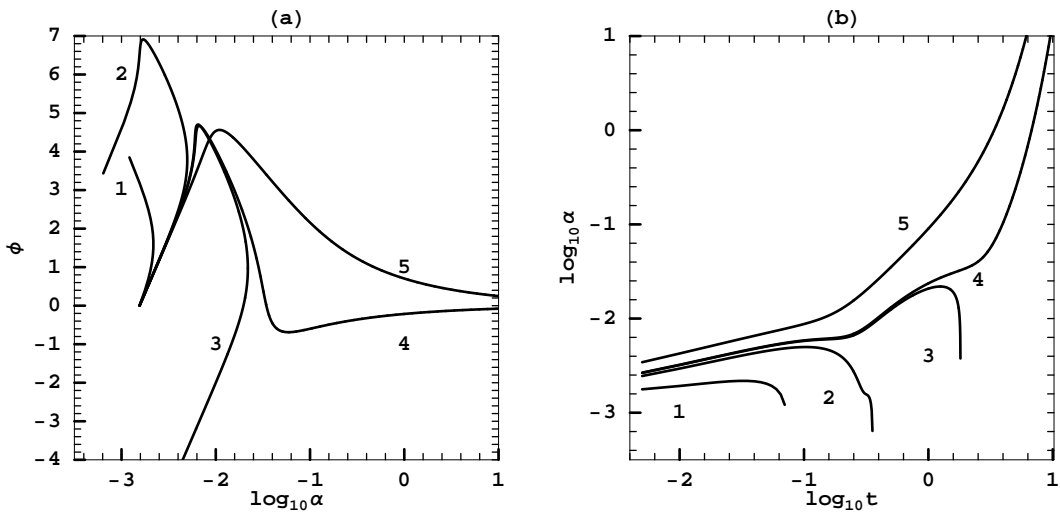}} 
  \caption{ (a) The scalar field $\phi$ vs. the logarithm of the scale factor 
  $ \log_{10} \alpha$. (b) The logarithm of the scale factor vs. the logarithm of 
  time $ \log_{10} t$ for 5 orbits in the semi-flat case ($ k = 10^{-3} $).
  }
         \label{Fighhaft}
 \end{figure*}
 \begin{figure*}
         \centerline{\includegraphics[width=35pc] {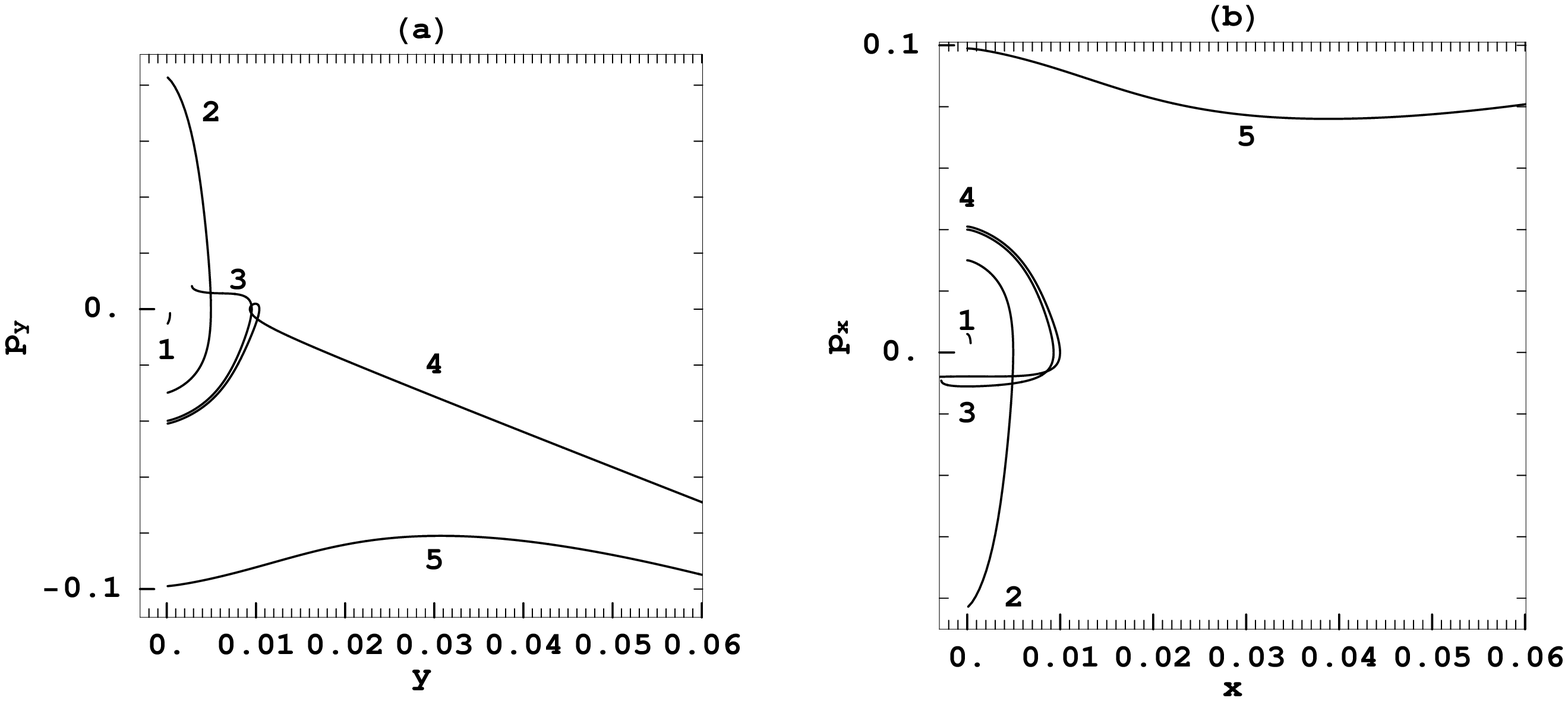}} 
  \caption{ (a) The projections of the orbits on the $y$, $p_y$ plane. (b)
   The projections of the orbits on the $x$, $p_x$ plane for 5 orbits in the semi-flat 
   case ($ k = 10^{-3} $).
  }
         \label{Fighhxy}
 \end{figure*}
 \begin{figure*}
         \centerline{\includegraphics[width=35pc] {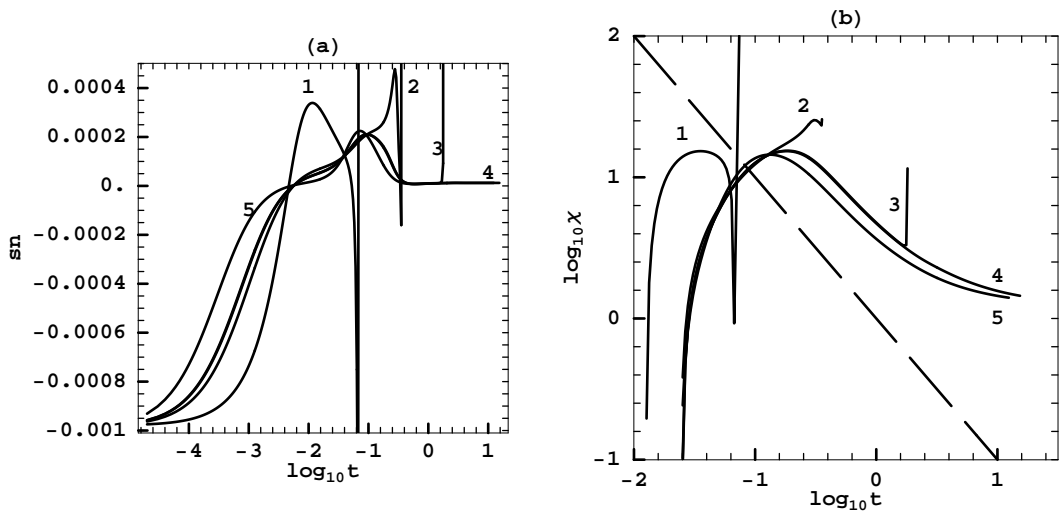}} 
  \caption{ (a) The stretching number ``sn'' versus the logarithm of time $\log_{10} t$. 
  (b) The logarithm of the finite time LCN, $log_{10} \chi$, versus the logarithm of 
  time $\log_{10} t$ for 5 orbits in the semi-flat case ($ k = 10^{-3} $). The dashed line in
  (b) represents the inclination $ -1 $ which is followed by the regular orbits. The initial,
  normalized to unity, deviation vector for all the orbits is $\frac{1}{2}(1,1,1,1)$ and the step
  of integration is $dt=10^{-5}$.
  }
         \label{Fighhch}
 \end{figure*}
 \begin{figure}
         \centerline{\includegraphics[width=20pc] {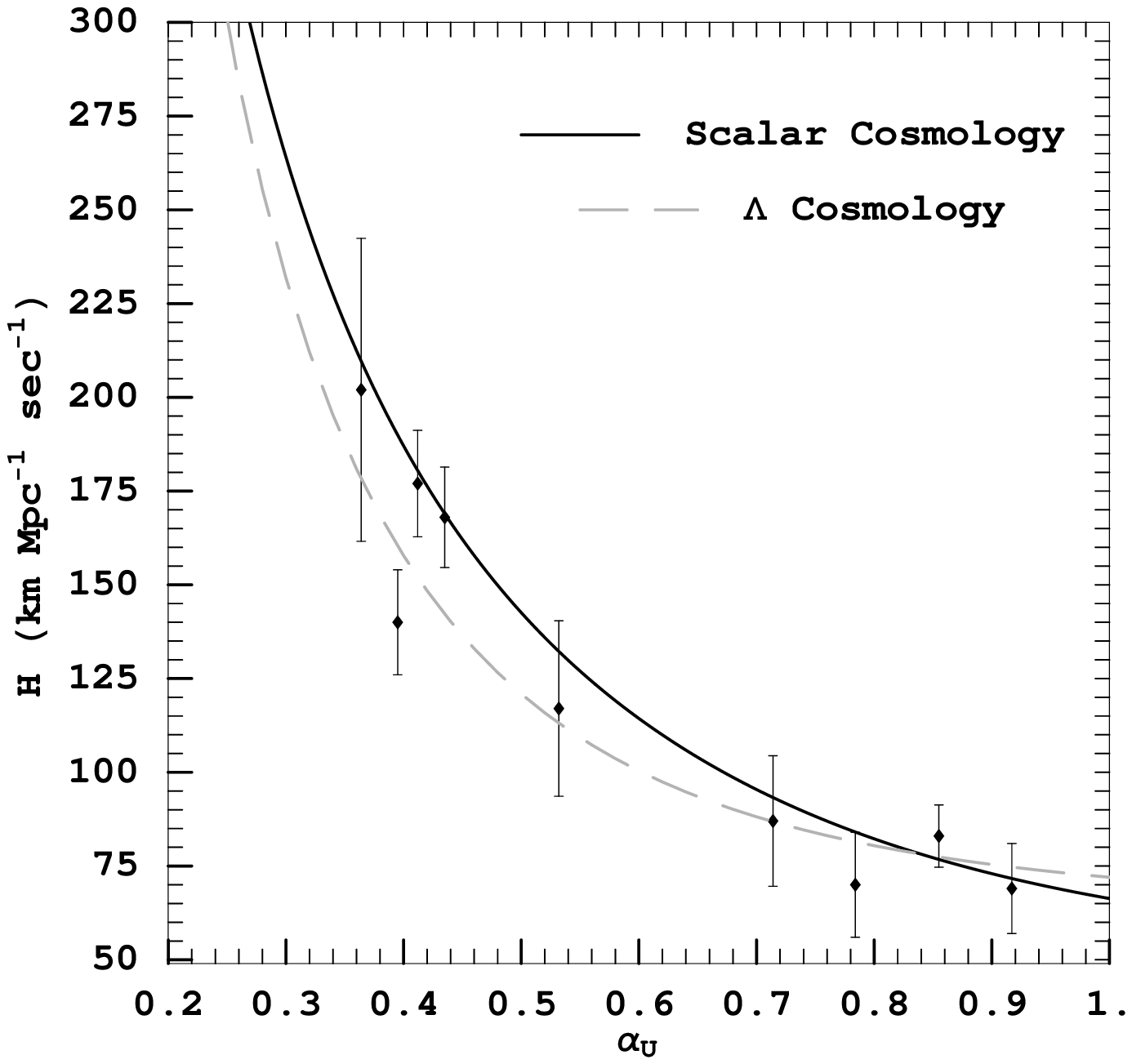}} 
  \caption{The Hubble parameter versus the scale factor of the 
           universe. 
           The points corresponds to the observational data.
           The gray dashed line represents the $\Lambda$ cosmology. The solid black line corresponds to the 
           the UDM model with $ k=10^{-3} $ and initial conditions 
           $x=5.6 \times 10^{-7},~y=0,~p_y=0.09 $. Note that the scale 
           factor scaled to unity at the present time. 
  }
         \label{Fighhub}
 \end{figure}
%
In the present section we investigate numerically the system of the RW coupled 
with a scalar field when the energy $ {\cal H} $ is zero. Our investigation refers
mainly to the case $ c_2 \geq 0 $ following the notations of \cite{Bertacca06}.
Note that in the appendix we present results for the case $c_2 < 0$.
If we introduce the noncanonical transformation ($k>0$) 
\begin{eqnarray}
      \label{noncant}
      x = k^{3/4} X & & y = k^{3/4} Y \nonumber \\
      p_x = k^{3/4} P_X & & p_y = k^{3/4} P_Y
\end{eqnarray}
then the Hamiltonian (\ref{Hamxy1}) gives:
\begin{eqnarray}
      \label{HamXY1}
      {\cal H} =\frac{k^{3/2}}{2} \times~~~~~~~~~~~~~~~~~~~~~~~~~~~~~~~~~~~~~~~~~~~ \nonumber \\
      \left[(P_{X}^2-\omega_1^2 X^2)-(P_{Y}^2-\omega_2^2 Y^2)-3^{4/3}(Y^2-X^2)^{1/3}\right]
\end{eqnarray}
The Hamiltonian (\ref{Hamxy2}) for $c_2<0$ transforms in the same way. The only difference
is that we have to change the sign of $\omega_1^2$ in (\ref{HamXY1}).
So as regards the parameter $ k $ for $ {\cal H} = 0$ we may consider only one value of k 
(here $k=10^{-3}$, hereafter called semi-flat case), because the transformation 
(\ref{noncant}) shows that any case of $ k $ can be transformed to another.
However, we must stress that the time rescales as $ k^{3/2}$ by the above transformation.
For other values of the curvature $k$ the value of $t$ must be multiplied by $ k^{3/2}$.
The reason we choose the semi-flat value of $k$ is that the analysis of the recent 
observations of the Cosmic Microwave Backround (CMB) anisotropies indicates that the 
spatial curvature of the universe takes a quite small positive value ($\Omega_k \approx 0.02$)
very close to zero \cite{Spergel07}. Our aim in this work is to investigate whether a small 
variation of the curvature around its nominal value ($k=0$) could affect the global 
dynamics of the universe. We do not present the cases where $ k < 0 $ because of the recent 
observational data. Furthermore, in the case of the
flat universe the two oscillators are uncoupled so the system is integrable.

In the case of two coupled hyperbolic oscillators we select $ c_1 = 1 $ and $ c_2 = 1 $
for our parameters, which means that the potential (\ref{potent}) is simply 
$U(\phi) = \cosh^2{( \sqrt{\frac{3}{8}} \phi)}+1$. Even such a simple parameterization
produces a variety of orbits. In order to show this variety when $ k=10^{-3} $
we choose five representative orbits. The initial conditions of these orbits are:
1) $ y = 0.0001 $, $ x = 0 $, $ p_x = 0.006 $,
2) $ y = 0.0001 $, $ x = 0 $, $ p_x = 0.03 $,
3) $ y = 0.0001 $, $ x = 0 $, $ p_x = 0.04 $,
4) $ y = 0.0001 $, $ x = 0 $, $ p_x = 0.041 $,
5) $ y = 0.0001 $, $ x = 0 $, $ p_x = 0.099 $. 
For all these orbits $ p_y $ is evaluated by using Eq. (\ref{Hamxy1}) with $ {\cal H} = 0 $ and
by choosing the negative sign in front of the square root  of $ p_y^2 $.

Figure \ref{Fighhaft}a shows the scalar field $\phi$ as a function of the logarithm of 
the scale factor $ \log_{10} \alpha$ for our five orbits and in the Fig. \ref{Fighhaft}b 
the logarithm of the scale factor as a function of the logarithm of time $ \log_{10} t$.
We see that the orbits 1, 2, 3 refer to universes that recollapse, therefore these orbits describe
examples of closed universes. On the other hand the orbits 4 and 5 refer to universes that expand 
to infinity, which means that even with a positive curvature we get cosmological models which 
expand for ever (see Ref. \cite{Krauss99}). It is obvious that as the curvature $k$ 
increases the orbits which describe closed universes will reach bigger values of the scale
factor before they collapse, which also suggests that they will collapse after a longer time.
It is interesting to mention that for those orbits which are in between 4 and 5 
the scale factor of the universe initially goes as $\propto t^{1/2}$, then for a certain 
interval of time as $\propto t^{2/3}$, and finally accelerates exponentially. In order to 
scale the dynamic time t of the system to the real time $t_U$ in Gyears, we fit 
our numerical results for the orbit 5 to the observational data. That way we find the relations 
$t_U\approx 6.1 t ~{\rm Gyrs}$ and $\alpha_U =3 \alpha$.

Although, the $(\alpha,\phi)$ coordinate system is the physically straight forward for the present
problem, the $(x,y)$ system gives a more clear insight to the problem. Thus in Fig. \ref{Fighhxy}a
we present the projections of the orbits on the $y$, $p_y$ plane, 
while in Fig. \ref{Fighhxy}b we give the projections of the orbits on the $x$, $p_x$ 
plane. In the $y$, $p_y$ plane the orbits 1, 2, 3 start from $ y=0 $ and finally return 
to $ y = 0 $. This is expected simply because these orbits model recollapsing universes 
$\alpha \rightarrow 0$. The orbits 4 and 5 describe expanding cases where $y$, like the scale
factor, tends to infinity and $ p_y \rightarrow - \infty $. The form of our orbits in the 
$x$, $ p_x$ plane are similar, except for the orbit 4 which, instead of giving $ x \rightarrow \infty $ 
returns to $ x=0 $ with $p_x \rightarrow 0$. This means that the orbit 4 is essentially driven by 
the $ y $, $ p_y $ oscillator which gains all the energy while the  $x$, $p_x$ motion ceases to exist.
Thus, the case of orbit 4 shows explicitely that the two coupled oscillators exchange energy. The
energy exchange takes place due to the coupling term. The question which arises now is if this exchange is done 
in a regular way or if it is done stochastically. To address this question we used the variational equations 
of (\ref{eqmot1}) to calculate the evolution of the stretching numbers and of the finite time LCN 
``$\chi$'' through time.

Doing so, in Fig. \ref{Fighhch} we present our results: in Fig.\ref{Fighhch}a is shown 
the stretching number (hereafter ``sn'') versus the logarithm of time $\log_{10} t$ and in Fig.\ref{Fighhch}b
the logarithm of $\chi$ ($\log_{10} \chi $) versus the logarithm of time $\log_{10} t$. The dashed  line 
in the Fig.\ref{Fighhch}b represents the inclination $ -1 $ which is followed by the regular orbits
as the LCN tends to 0. This dashed line helps us to separate the regular from the chaotic orbits, because
the regular orbits always evolve along a parallel line, though usually with some variations around the line,
while the chaotic orbits change after a certain time point their inclination from -1 to 0 as the $\chi$
tends to a finite value. For all the considered orbits Fig \ref{Fighhch}a the ``sn'' stays for most of the time 
above 0 and this indicates that for all five orbits $\chi$ should be positive, therefore there is a strong
indication that these orbits are chaotic. This conclusion seems to agree with the results which we see also
for the orbits 4 and 5 in Fig. \ref{Fighhch}b, but the orbits 1, 2 and 3 are not so clear cases because they
recollapse. In particular, the expanding to infinity orbits 4 and 5 appear to be chaotic. Their $\chi$ 
(see Fig. \ref{Fighhch}b) tends to a finite nonzero value $ \approx \sqrt{2}$ and the ``sn''
is on the average positive. This finite $\chi$ value is the eigenvalue  $\sqrt{c_1+c_2}$
(\ref{eig2}) of the unstable manifold, which emanates from the equilibrium point, for $c_{1}=c_{2}=1$.
Notice that this eigenvalue does not depend on the curvature, hence for any positive curvature in 
the case under study we expect that the chaotic orbits will have the same $LCN$. This expectation was numerically
attested for several orders of magnitude of the curvature's value in the case $c_{1}=c_{2}=1$. 
This is true if the orbits do not recollapse. If the orbits recollapse before the $ \chi $ indicator can decide
whether they are chaotic, then the ``sn'' is the only indicator that can suggest the chaoticity 
of these orbits. In the case of the orbits 1, 2 and 3 (Fig. \ref{Fighhch}a) the ``sn'' indeed indicates 
that these orbits are chaotic.

The recollapse for the orbits 1,2 and 3 in Figs. \ref{Fighhch}a,b can be observed by an abrupt change of 
``sn'' and $\chi$ indicators respectively. At the recollapse both indicators' values start to fall, abruptly 
for orbits 1 and 2, marginally for orbit 3, before they steeply rise and become infinite. 
The fall of the indicators at the recollapse is explained by the fact that when an orbit falls to the 
anomaly $\alpha = 0$ the position variables tend to stabilize, thus the deviations of the position 
variables shrink and the deviation vector's norm gets smaller. As long as the value of the deviation 
vector's norm decreases, the ``sn'' becomes more and more negative and therefore the $ \chi $ value decreases.
However, as the moving point gets nearer to the anomaly the momentum variables start to grow and their 
deviations expand. This means that the deviation vector's norm increases, thus the ``sn'' increases and the
$ \chi $ value rises. Because the momentum variables grow abruptly the rise of ``sn'' and $\chi$ is steep.

We should state here that the anomalies produced by the 
recollapse could be avoided by a logarithmic 
transformation of the time, as done in the mixmaster problem (for a review see \cite{Contop02}). 
However, we do not discuss here the usefulness of such a transformation, because the recent 
high quality cosmological data show that the universe is expanding accelerated to infinity 
and does not recollapse.

We emphasize that the final value of $\chi$ does not depend on the curvature's value $k$. Hence
cosmological models for any positive value of $k$ reveal a chaotic behavior in contrast with 
the pure flat models ($k=0$). As a special case, semiflat models ($k=10^{-3}$) with 
the characteristics close to orbits 4 and 5 predict an overall dynamics which 
is close to the $\Lambda$ cosmology 
\footnote{In the $\Lambda$ cosmology the scale factor of the universe is 
$\alpha(t)=(\Omega_M/\Omega_{\Lambda})^{1/3} {\rm sinh}^{2/3}(
3 H_0 \sqrt{\Omega_{\Lambda}}
 t/2)$ and the Hubble parameter is 
$H(\alpha)=H_{0}\sqrt{\Omega_{\rm m}\alpha^{-3}+\Omega_{\Lambda}}$.
In this work we use $H_0\simeq 72$Kms$^{-1}$Mpc$^{-1}$ 
\cite{Freedman01} and 
$\Omega_{\rm m}=1-\Omega_{\Lambda}\simeq 0.26$ \cite{Spergel07}.},
despite the fact that for these UDM cosmological paradigms there is a strong indication
of a chaotic behavior! 
Indeed, orbit 5 starts with a state which is similar to the radiation epoch
($\alpha \propto t^{1/2}$), then evolves to the matter epoch ($\alpha \propto t^{2/3}$) and 
finishes with the exponential growth 
of the universe

 \begin{figure}
         \centerline{\includegraphics[width=17pc] {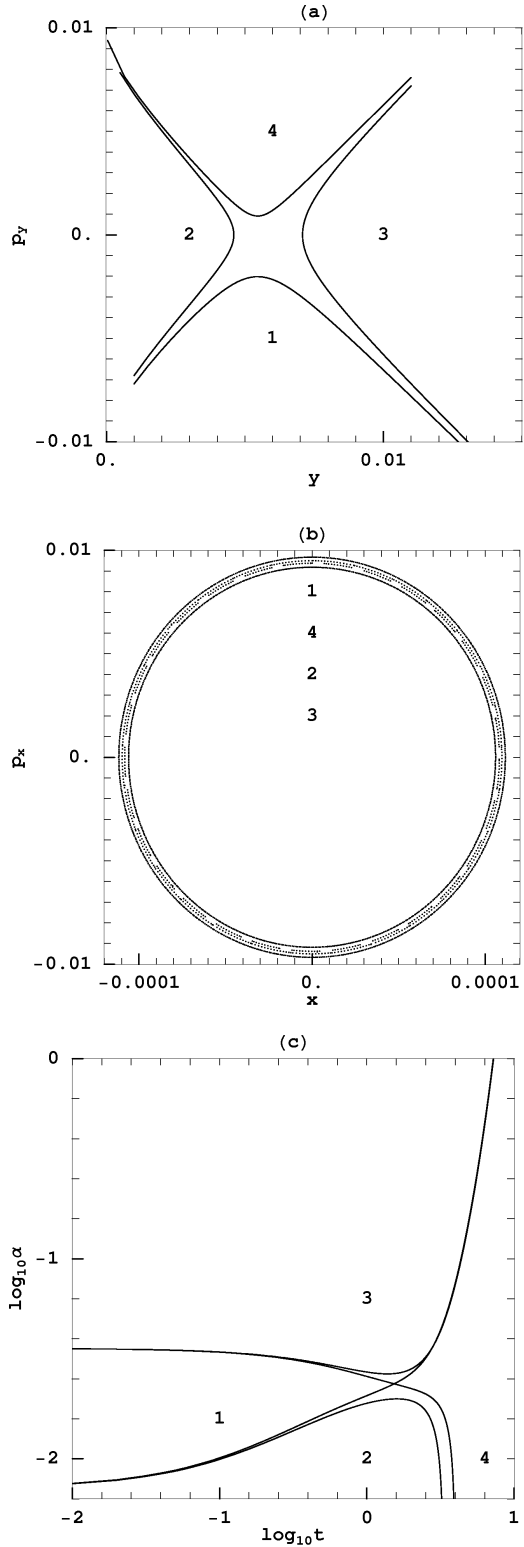}} 
  \caption{ (a) The Poincare\'{e} section $ x = 0 $ with $ p_x > 0 $. (b) The orbits' 
  projections on the $x$, $p_x$ plane for 4 orbits. (c) The development of $ \log_{10}\alpha $ 
  as a function of the logarithm of time $ \log_{10} t $ with $ k = 10^{-3} $.
  }
  \label{Fighe}
 \end{figure}

On the other hand, it is interesting to mention that 
the interplay between the values of the initial conditions  
could yield semiflat UDM cosmological models 
which predict well the observed evolution of the Hubble parameter. 
As an example, in  Fig. 4 we plot the measured Hubble
parameter together with a UDM scenario (solid line)
which is formed by the following 
initial conditions: $x=5.6 \times 10^{-7},~y=0,~p_y=0.09$.
It is obvious that this model reproduces, 
even better than the $\Lambda$ cosmology (dashed line), 
the observational data. Note that the chaotic behavior of this UDM paradigm 
is the same with that found in the case of orbit 5.
In a subsequent paper we shall investigate statistically the parameter space
for models with $k \geq 0$ which fit best the available observational data.

The above result is novel because within the framework of the scalar field 
cosmology we can find semiflat models with specific initial conditions which 
reveal chaos and at the same time provide an evolution of the Hubble parameter
close to the observed one. In any case, we expect that the exact value of the curvature 
will be finally measured by the next generation of the CMB data (PLANCK observatory)
and therefore, we cannot preclude observational surprises
regarding the curvature of the universe. 

\section{Conclusions}

In this work we investigate the dynamics of the closed scalar field FRW cosmological models in the 
framework of the so called {\it Unified Dark Matter} scenario. We find that for the closed geometry
there is a strong indication of chaos in agreement with previous studies and that the $LCN$ of these
models is independent from the curvature. In particular, we find that there are semiflat cosmological
models with specific initial conditions in which there is a clear indication for a chaotic behavior 
and at the same time the corresponding dynamics is almost the same as in the $\Lambda$ cosmology.
We verify this by combaning the measured evolution of the Hubble parameter with that
expected in the scalar field cosmology (with specific initial conditions) and we find a very good agreement. 
If that is the case, then the chaotic fields may provide an alternative theory for the solution 
of the cosmological coincidence problem.

\begin{acknowledgments}
G. Lukes-Gerakopoulos was supported by the Greek Foundation of State Scholarships (IKY).
\end{acknowledgments}

\appendix

\section{Case $ c_2 < 0 $}

In this kind of dynamical studies, it is well known (for details see \cite{Contop02}) 
that one of the best procedures to find chaos is based on the so called Poincar\'{e} sections.
Unfortunately, in the case of  $ c_2 \geq 0 $ the Poincar\'{e} sections are not applicable, 
due to the fact that the orbits do not exhibit recurrences. On the other hand, for $ c_2 < 0 $ 
Poincar\'{e} sections can be obtained because of the 
elliptic oscillator on $x$, $p_x$ plane and that only if the frequency of the elliptic oscillator
is much higher than the frequency of the hyperbolic one. Hence in order to achieve the latter 
we use here $ c_1 = 10^4+1 $ and $ c_2 = 1-10^4 $. We mention that the case $c_2<0 $ is mainly
of mathematical interest, because, as shown in Ref. \cite{Bertacca06}, $c_2$ is meaningful
only if it is not negative.

In Fig. \ref{Fighe}a are shown Poincar\'{e} sections for $k=10^{-3}$.
The surface of section is for $ x = 0 $ and $ p_x>0 $. From these Poincar\'{e}
sections we conclude that the present dynamical system's phase space is separated into
4 subspaces with specific directions of flow. Each subspace has a typical type of orbits,
therefore we need only 4 appropriate initial conditions in order to 
represent the whole section's phase space. The initial conditions we choose are
 1) $ y = 0.001 $, $ p_y = -0.0072 $, $ x = 0 $, 
 2) $ y = 0.001 $, $ p_y = -0.0068 $, $ x = 0 $, 
 3) $ y = 0.011 $, $ p_y = 0.0072 $, $ x = 0 $, 
 4) $ y = 0.011 $, $ p_y = 0.0076 $, $ x = 0 $.
Note that in this subsection $ p_x $ is calculated from Eq. (\ref{Hamxy2})
with $ {\cal H} = 0 $ and by choosing the positive sign in front of the square root of $p_x^2$.
The direction of the flow for orbits of type:
\begin{enumerate}
 \item is along the $ y $ direction.
 \item is along the $ p_y $ direction.
 \item is along the negative $ y $ direction.
 \item is along the negative $ p_y $ direction.
\end{enumerate}
The equilibrium point in the Poincar\'{e} section is at the center of the 
cross formed by the representative orbits, as given by Eqs. (\ref{eqp2a}). 
The elliptic oscillator can be seen in Fig. \ref{Fighe}b where we give 
the projection of each orbit on the $x$, $p_x$ plane. As expected, the 
orbits on the $ x $, $ p_x$ projection move almost on ellipses whose axes have a ratio
$\frac{p_x}{x} \approx \frac{1}{100}$. The numbers in Fig. \ref{Fighe}b written on 
the $ x = 0 $ axis indicates how the four orbits are ordered according to their
$ p_x $ axis length. We mention that the length of each axis varies as a function of time.

From the dynamical point of view, which we get from Figs. \ref{Fighe}a,b, we now move on
to the more physical point of view of Fig. \ref{Fighe}c. In this figure we give
the development of $ \log_{10}\alpha $ as a function of the logarithm of time $ \log_{10}t $. 
In these figures we see orbits of the following variety. Models:
\begin{enumerate}
 \item  describe a cosmological model which starts from $\alpha \rightarrow 0 $ and ends
 with $\alpha \rightarrow \infty $ after an inflection point.
 \item  describe a cosmological model which starts from $\alpha \rightarrow 0 $ and collapses.
 \item  describe a cosmological model which starts from a finite $\alpha $ and ends
 with $\alpha \rightarrow \infty $.
\item  describe a cosmological model which starts from a finite $\alpha $ and collapses
 after an inflection point. This infection point represent the passing of the scale factor
 from $ \ddot{\alpha} > 0 $ to $ \ddot{\alpha}< 0 $
\end{enumerate}
Models of type 1 are the only ones that could be compatible with the standing cosmology, 
but the scale factor at first increases very slowly with t and in the end it increases
superexponentially. As for the chaoticity of the orbits in the $c_2 < 0$ case, the indexes 
$\chi$ and ``sn'' indicate that these orbits are chaotic.

We studied also some cases with $ {\cal H} \neq 0 $ for both $ c_2 \geq 0$ and $ c_2 < 0 $. 
The whole phase space in all cases is dependent on the position of the equilibrium point, because 
the asymptotic manifolds emanating from the equilibrium point determines the flow in the phase space 
and therefore determines which models will collapse and which shall expand to infinity. 
As can be seen from Eqs. (\ref{eqp1a}),(\ref{eqp2a}) the equilibrium point is independent 
from $ {\cal H} $, so the main characteristics of the $ {\cal H} \neq 0 $ cases are similar 
to those with $ {\cal H} = 0$.

\end{document}